\documentstyle[11pt]{article}

\newcommand{\sect}[1]{\setcounter{equation}{0}\section{#1}}
\newcommand{\subsect}[1]{\subsection{#1}}
\newcommand{\subsubsect}[1]{\subsubsection{#1}}
\renewcommand{\theequation}{\arabic{section}.\arabic{equation}}

\def\be{\begin{equation}}
\def\ee{\end{equation}}
\def\bea{\begin{eqnarray}}
\def\eea{\end{eqnarray}}

\def\co{\Delta}  
\def\Dco{\Delta}

\def\k{\kappa} 
\def\t{\theta}

\def\J{{\tilde J}} 
\def\j{{\tilde \sigma}} 
\def\s{{\sigma}} 

\def\>#1{{\bf #1}}                 
\def\1{\'{\i}}                           
\def\R{\rm I\kern-.2em R} 
\def\C{\rm I\kern-.5em C}

\def\tfrac#1#2{ {\scriptstyle { \frac {#1}{#2}}}}         
\def\pois#1#2{\left\{ {#1},{#2} \right\}}         
\def\ponm#1#2{\left [ {#1},{#2} \right\}}         
\def\conm#1#2{\left [ {#1},{#2} \right ]}         

\def\osc{h_4}

\def\aa{N} 
\def\ap{{A_+}} 
\def\am{{A_-}}
\def\bb{M}

\begin{document}
\thispagestyle{empty}
\vspace{2.5cm}

 \

\begin{center} {\LARGE{\bf{A systematic construction of 
completely}}} 
  
 {\LARGE{\bf{integrable Hamiltonians from coalgebras}}} 

\vspace{1.4cm} 

ANGEL BALLESTEROS\vspace{.2cm}\\
{\it Departamento de F\1sica, Universidad de Burgos\\
Pza. Misael Ba\~nuelos s.n., 09001-Burgos, Spain} 
\vspace{.3cm}

\vspace{0.4cm} 

ORLANDO RAGNISCO\vspace{.2cm}\\
{\it Dipartimento di Fisica, Terza Universit\'a di Roma\\
Via Vasca Navale 84, Roma, Italy} 
\vspace{.3cm}

\end{center} 
  
\bigskip

\begin{abstract} 

A universal algorithm to construct $N$-particle (classical and quantum)
completely integrable Hamiltonian systems from representations of coalgebras
with Casimir element is presented. In particular, this construction shows
that quantum deformations can be interpreted as generating structures for
integrable deformations of Hamiltonian systems with coalgebra symmetry. In
order to illustrate this general method, the $so(2,1)$ algebra and the
oscillator algebra $h_4$ are used to derive new classical integrable systems
including a generalization of Gaudin-Calogero systems and oscillator chains.
Quantum deformations are then used to obtain some explicit integrable
deformations of the previous long-range interacting systems and a
(non-coboundary) deformation of the $(1+1)$ Poincar\'e algebra is shown to
provide a new Ruijsenaars-Schneider-like Hamiltonian. 

\end{abstract} 

\bigskip
\newpage


\sect{Introduction}

It is well-known that quantum groups appeared in the 
context of Quantum Inverse Scattering Method as a new kind of symmetries
linked to the integrability of some quantum models constructed in Lax form 
(see \cite{KBI,GRAS,FRT}). Quantum algebras and groups
are related by duality \cite{Dr} and, in the previous context, the concept
of ``quantum algebra invariance" expresses the commutativity of a given
Hamiltonian with respect to the generators of a certain quantum algebra.
Since their introduction, the construction and analysis of quantum group
invariant integrable models has concentrated many efforts (see
\cite{PS,KS,ACF,Chang,Foerster} and references therein) and a great amount
of literature has been also devoted to quantum
group theory (see, for instance, \cite{CP}). 

From an abstract mathematical point of view,
two ideas were emphasized as a consequence of these developments: the
relevance of deformations (in the sense of \cite{Gerst} and \cite{BFFLS})
and the concept of Hopf algebra \cite{Hopf}. In
particular, quantum algebras are just defined as Hopf algebra deformations
of usual universal enveloping Lie algebras. On the other
hand, although quantum semisimple algebras were the ones initially linked
to integrable models, the construction of quantum deformations of 
non-semisimple Lie algebras has been also succesfully explored by using
different methods (see \cite{CGST,Osc}). 

This paper establishes a
general and constructive connection between Hopf algebras and integrability
that can be stated as follows: Given {\it any} coalgebra $(A,\co)$ with
Casimir element $C$, each of its representations gives rise to a family of
completely integrable Hamiltonians $H^{(N)}$ with an arbitrary number $N$
of degrees of freedom. We provide a constructive proof of this statement
that contains the explicit definition of such Hamiltonians and their
integrals of motion. Moreover, both classical and quantum mechanical
systems can be obtained from the same $(A,\co)$, provided we endow this
coalgebra with a suitable additional structure (that will be either a
Poisson bracket or a non-commutative product on $A$, respectively). Note
that, instead of Hopf algebra structures, our construction makes use of the
more general term of coalgebra (neither the counit nor the antipode mappigs
will be explicitly used).  

It is important to emphasize that the validity of this general procedure by
no means depends on the explicit form of $\co$ (i.e., on whether the
coalgebra $(A,\co)$ is deformed or not). This fact is crucial in order to
clarify the significance of quantum algebras (and groups) in our framework:
they are ``only" a particular class of coalgebras that can be used to
construct systematically integrable systems. But the specific feature of
such systems will be that they are {\it integrable deformations} of the
ones obtained by the same method when we start from the corresponding
non-deformed coalgebra. Moreover, usual Lie algebras are always endowed
with a coalgebra structure, and we shall see that many interesting
coalgebra-induced systems can be derived from them without making use of
any deformation machinery.  In this way, a new general application (and
intrinsically different from the usual ones \cite{OP,Per})  of Lie algebras
in the field of integrable systems is presented. At this point, we would
like to mention that this result concerning non-deformed Lie coalgebras was
already proven in \cite{BCR}, and it can be also extracted from
\cite{KarAlg}, but without explicit mention of the underlying coalgebra
structure.

In the next Section the basics of Hopf algebras are revisited and the
definition and properties of Poisson coalgebras presented.
Realizations of Poisson coalgebras on canonical coordinates are introduced
and coupled with the coproduct map in order to get two-particle
representations. Section 3 is devoted to the
construction of a family of $N=2$ integrable systems from a $so(2,1)$
Poisson coalgebra structure. An integrable deformation of this family is
afterwards obtained by making use of the (standard) quantum deformation of
this algebra. This two-dimensional example contains the seminal ideas of
our construction, that need a proper generalization in order to reach the
full
$N$-dimensional case. The mathematical improvements needed to succeed in
such a general scheme are presented in Section 4, that contains an analysis
of the usual definition of the $N$-th coproduct map $\co^{(N)}:A\rightarrow
A\otimes A\otimes\dots^{N)}\otimes A$ in terms of a recurrence relation
that starts with the second order coproduct $\co\equiv\co^{(2)}$. It turns
out that it is possible to rewrite $\co^{(N)}$ in a different way that is
much more convenient for our purposes.  

Section 5 introduces the general constructive result, valid for both
classical and quantum mechanical systems: the $N$-th coproduct of any
(smooth) function of the generators of a coalgebra defines an integrable
Hamiltonian whose constants of motion in involution are given by the $m$-th
coproducts of the Casimir element $C$, with $m=1,\dots,N$. Functional
independence among the constants is guaranteed by construction. Some
comments concerning the specific features of both the classical and the
quantum mechanical cases are included.

In order to show the direct applicability of these results, Section 6 
includes various examples based on phase-space realizations of coalgebras
(although the quantization of some of them is not difficult, a careful
treatment of some quantum mechanical examples will be presented in a
forthcoming paper). The first one makes use of the classical $so(2,1)$
Poisson coalgebra in order to construct a new multiparameter generalization
of an integrable system that has been recently introduced by Calogero
\cite{Cal} and whose coalgebra symmetry is manifestly extracted. The second
non-deformed example is provided by the (primitive) coalgebra linked to the
(non-semisimple) oscillator algebra $h_4$, that leads to a straighforward
proof of the integrability of a system of coupled oscillators firstly given
in \cite{CalOsc}. Afterwards, the fact that quantum algebras can be
interpreted as the generating objects of integrable deformations is
illustrated by using the standard quantum deformation of $so(2,1)$ to get
-through suitable Poisson realizations of this quantum algebra- an
integrable deformation of the previous $so(2,1)$ family. Finally, another
interesting example of quantum algebra induced integrable system is
provided by the Vaksman-Korogodskii deformation \cite{VK} of the (1+1)
Poincar\'e algebra, that gives rise to a Ruijsenaars-Schneider like
integrable Hamiltonian \cite{RS}. It is interesting to note that, in this
case, the system defined through the non-deformed (1+1) Poincar\'e algebra
is quite trivial; consequently, the quantum deformation seems to be
sometimes essential in order to produce a dynamically relevant Hamiltonian.

In Section 7 a deeper insight into the $so(2,1)$ models is presented
by precluding the use of canonical realizations and working with classical
``angular momentum variables". In this way, the long-range nature of the
interaction of these models is clearly appreciated. Under this realization,
the non-deformed coalgebra gives rise to the hyperbolic XXX Gaudin magnet
\cite{Gau}, and the integrable deformation linked to $U_z(so(2,1))$ is
translated into physical terms as the introduction of a variable range
exchange \cite{Inoz} within the Gaudin Hamiltonian. Finally, the paper is
closed by some remarks concerning open questions and future developments.

\sect{Coalgebras and Poisson realizations}

\subsect{Hopf algebras}

A {\it Hopf algebra} is a (unital, associative) algebra $(A,\cdot)$ endowed
with two homomorphisms called coproduct $(\Delta : A\longrightarrow
A\otimes A )$ and counit $(\epsilon : A\longrightarrow \C)$, as well as an
antihomomorphism (the antipode $\gamma : A\longrightarrow A$) such that,
$\forall a\ \! \in A$:
\bea
&(id\otimes\co)\co (a)=(\co\otimes id)\co (a),  
\label{co}\\
&(id\otimes\epsilon)\co (a)=(\epsilon\otimes id)\co (a)= a, 
\label{coun}\\
&m((id\otimes \gamma)\co (a))=m((\gamma \otimes id)\co (a))=
\epsilon (a) 1, 
\label{ant}
\eea
where $m$ is the usual multiplication mapping $m(a\otimes b)=a\cdot b$.
This notion was introduced by Hopf \cite{Hopf} in a
cohomological context but, as we shall see, it expresses a basic idea in
many-body problems and it is often implicitly used. The aim of this paper
is just to make more explicit its physical significance, that is basically
concentrated within the coproduct $\Delta$. In fact, hereafter we shall
deal mainly with coalgebras, i.e., algebras endowed with a coassociative
(\ref{co}) coproduct $\Delta$.

For our purposes, the most interesting example of coalgebra is provided by 
the universal enveloping algebra $U(g)$ of a Lie algebra $g$ with generators
$X_i$. The algebra $U(g)$ can be endowed with a Hopf algebra
structure by defining: \bea &\co(X_i)=1\otimes X_i + X_i\otimes 1,\quad
\co(1)=1\otimes 1,\cr &\epsilon(X_i)=0,\quad \epsilon(1)=1,\cr
&\gamma(X_i)=-X_i, \quad \gamma(1)=1. \label{uea}
\eea
These maps acting on the generators of $g$ are straightforwardly extended
to any monomial in $U(g)$ by means of the homomorphism condition
$\co(X\cdot Y)=\co(X)\cdot \co(Y)$. In general, an element $Y$ of a Hopf
algebra such that $\co(Y)=1\otimes Y + Y\otimes 1$ is called {\it
primitive}, and Friedrichs' Theorem ensures that, in $U(g)$, the only
primitive elements are the generators $X_i$ \cite{Post}. On the other hand,
the homomorphism condition implies the compatibility of the coproduct
$\Delta$ with the Lie bracket
\be
\conm{\co(X_i)}{\co(X_j)}_{A\otimes A}=\co(\conm{X_i}{X_j}_A),\qquad \forall
X_i,X_j\in g. \label{hom}
\ee

From a physical point of view, if $g$ is the algebra of observables of some
one-particle physical system, the coproduct in (\ref{uea}) is just the
usual definition of ``total" quantum observables for the two-particle
system.

In this context, quantum algebras are just coalgebra deformations of $U(g)$: 
a deformed, but coassociative, coproduct is defined and a set of
(possibly deformed) commutation rules can be found in such a way that the
compatibility condition (\ref{hom}) is recovered. The whole ``quantum"
structure depends on (perhaps more than one) deformation parameters and the
non-deformed coalgebra (\ref{uea}) is recovered when all the parameters
vanish. A well known example is the standard (Drinfel'd-Jimbo
\cite{Dr,Ji}) deformation of $U(so(2,1))$ with deformed coproduct
\bea
&& \Dco(\J_2) =1 \otimes \J_2 + \J_2\otimes 1,\cr  
&&  \Dco(\J_1) =e^{-\tfrac{z}{2}\J_2} \otimes \J_1 + \J_1\otimes
e^{\tfrac{z}{2}\J_2}; \label{lb} \\
&&  \Dco(\J_3) =e^{-\tfrac{z}{2}\J_2} \otimes \J_3 + \J_3\otimes
e^{\tfrac{z}{2}\J_2};\nonumber
\eea
and deformed commutation rules compatible with (\ref{lb})
\be
\conm{\J_2}{\J_1}=\J_3,\quad \conm{\J_2}{\J_3}=-\J_1,\quad 
\conm{\J_3}{\J_1}=\frac{\sinh (z\J_2)}{z}.
\label{lc} 
\ee
Another important object is essential for our purposes: the existence of a
deformed Casimir that commutes with all the generators of the quantum
algebra and, in this case, reads
\be
C_z(\J_1,\J_2,\J_3)=\left( 2\,\frac{\sinh (\tfrac{z}{2} \J_2)}{z} 
\right)^2 -
\J_1^2 - \J_3^2. \label{le}
\ee 
As we shall see, both deformed and non-deformed Casimir elements will be
the keystone of the integrability properties of the systems induced from
their respective coalgebras.

\subsect{Poisson coalgebras and canonical realizations}

In general, a {\it Poisson algebra} $P$ is a vector space endowed with a
commutative multiplication and a Lie bracket $\pois{\,}{\,}$ that
induces a derivation on the algebra in the form
\be
\pois{a}{b\,c}=\pois{a}{b}\,c + b\,\pois{a}{c},\qquad \forall
a,b,c\in P. 
\label{der}
\ee
If $P$ and $Q$ are Poisson algebras, we can define the following
Poisson structure on
$P\otimes Q$:
\be
\pois{a\otimes b}{c\otimes d}_{P\otimes Q}:=\pois{a}{c}_P\otimes
b\,d + a\, c\otimes \pois{b}{d}_Q.
\label{stc}
\ee

We shall say that $(A,\co)$ is a {\it
Poisson coalgebra} if $A$ is a Poisson algebra and the
coproduct $\co$ is a Poisson algebra homomorphism between $A$ and
$A\otimes A$:
\be
\pois{\co_A(a)}{\co_A(b)}_{A\otimes
A}=\co(\pois{a}{b}_A),\qquad \forall a,b\in A.
\label{std}
\ee
Obviously, given any Lie algebra $g$ a Poisson coalgebra can be obtained by
defining a Poisson bracket by means of the bivector
\be
\Lambda=c_{ij}^k\, x_k \, \partial_{x_i}\wedge\partial_{x_j},
\label{sieb}
\ee
where the $x$ are local coordinates on a certain manifold linked to the
generators of $g$ and $c_{ij}^k$ is the structure tensor for $g$. It is
immediate to check that the coproduct (\ref{uea}) is a Poisson map if the
Poisson bracket on the tensor product space is defined by (\ref{stc}). 
Quantum deformations can be also realized as Poisson coalgebras in this way: a
natural Poisson coalgebra linked to $U_z(so(2,1))$ is given by the bivector
\be
\Lambda=\j_3 \, \partial_{\j_2}\wedge\partial_{\j_1} - 
\j_1 \, \partial_{\j_2}\wedge\partial_{\j_3} + 
\frac{\sinh (z\,\j_2)}{z} \, \partial_{\j_3}\wedge\partial_{\j_1} ,
\label{biq}
\ee
and the coproduct (\ref{lb}) where the quantum algebra generators are
replaced by their corresponding local coordinates $\j_i$ on $\R^3$.
Obviously, the Poisson structure (\ref{biq}) wil be non-degenerate on the
symplectic leaf defined by
\be
\left( 2\,\frac{\sinh (\tfrac{z}{2} \j_2)}{z}\right)^2 -
\j_1^2 - \j_3^2 = c_z.
\label{sym}
\ee 

On the other hand, the connection between a Lie algebra and a
one-particle system can be made explicit by considering that $g$ is
realized by means of smooth functions on the one--particle phase space
$\R^2$ with local coordinates $(p,q)$
\be
D(X_i)=X_i(p,q).
\label{rea}
\ee
This means that, under the
``canonical" Poisson bracket
\be
\pois{f}{h}=\frac{\partial f}{\partial q}
\frac{\partial h}{\partial p} - \frac{\partial h}{\partial q}
\frac{\partial f}{\partial p},\quad f,h\in C^\infty(p,q),
\label{si}
\ee
the ``generators" (\ref{rea}) close the initial Lie algebra:
\be
\pois{X_i(p,q)}{X_j(p,q)}= c_{ij}^k\,X_k(p,q).
\label{lie}
\ee
Two different one-particle realizations (\ref{rea}) will be equivalent if
there exists a canonical transformation that maps one into the other. A
simple example is given by the following one-particle realization of the
Poisson coalgebra linked to $so(2,1)$:
\be
D(J_2)=p,\qquad
D(J_1)=p\,\cos q,\qquad
D(J_3)=p\,\sin q.
\label{hb}
\ee
This realization (that leads to a vanishing Casimir function) can be easily
deformed:
\be
D_z(\J_2)=p,\qquad
D_z(\J_1)=2\,\frac{\sinh (\tfrac{z}{2} p)}{z}\,\cos q,\qquad
D_z(\J_3)=2\,\frac{\sinh (\tfrac{z}{2} p)}{z}\,\sin q,
\label{ld}
\ee
These phase-space functions close a quantum $so(2,1)$ algebra (\ref{lc})
under the canonical Poisson bracket (\ref{si}).

Now, the essential feature of a Poisson coalgebra becomes evident: if we
represent $A\otimes A$ by using two copies of
(\ref{rea}), the functions
$\co(X_i)(q_1,q_2,p_1,p_2)$ (we use the notation $p\otimes
1\equiv p_1$, $1\otimes p\equiv p_2$, and so on) define the same Lie algebra
$g$
\be
\pois{\co(X_i)}{\co(X_j)}_{A\otimes
A}=\co (\pois{X_i}{X_j}_A)=c_{ij}^k\,\co(X_k),\qquad \forall X_i,X_j,
\label{sid}
\ee
with respect to a bracket (\ref{sid}) given by
\be
\pois{f}{h}=\sum_{i=1}^{2}{\left(\frac{\partial f}{\partial q_i}
\frac{\partial h}{\partial p_i} - \frac{\partial h}{\partial q_i}
\frac{\partial f}{\partial p_i}\right)}.
\label{sie}
\ee
In particular, (\ref{sie}) leads to the Poisson bracket (\ref{stc})
provided we have chosen $f=a(q_1,p_1)\,b(q_2,p_2)$ and
$h=c(q_1,p_1)\,d(q_2,p_2)$.

In the case of $so(2,1)$, this coalgebra
property means that the following two-particle functions defined through
the coproduct (\ref{uea}) and the realization (\ref{hb})
\bea
&& f_1(\vec{q},\vec{p})=(D\otimes D)(\co(J_1))=p_1\,\cos q_1 + p_2\,\cos
q_2,\nonumber\\ 
&& f_2(\vec{q},\vec{p})=(D\otimes
D)(\co(J_2))=p_1+p_2,\label{dosclas}\\
&& f_3(\vec{q},\vec{p})=(D\otimes
D)(\co(J_3))=p_1\,\sin q_1 + p_2\,\sin q_2, \nonumber
\eea
close the $so(2,1)$ algebra. The deformed construction is also immediate:
from (\ref{lb}) and (\ref{ld}) we obtain the functions
\bea
&& f_1^z(\vec{q},\vec{p})=(D_z\otimes D_z)(\co(\J_1))=2\,
\frac{\sinh (\tfrac{z}{2} p_1)}{z}
\,\cos q_1\,e^{\frac{z}{2}\,p_2} + e^{-\frac{z}{2}\,p_1}\,2\,\frac{\sinh
(\tfrac{z}{2} p_2)}{z}\,\cos q_2,\nonumber\\
&& f_2^z(\vec{q},\vec{p})=(D_z\otimes D_z)(\co(\J_2))=p_1+p_2,
\label{dosq}\\
&& f_3^z(\vec{q},\vec{p})=(D_z\otimes D_z)(\co(\J_3))=2\,
\frac{\sinh (\tfrac{z}{2} p_1)}{z}
\,\sin q_1\,e^{\frac{z}{2}\,p_2} +
 e^{-\frac{z}{2}\,p_1}\,2\,\frac{\sinh (\tfrac{z}{2}
p_2)}{z}\,\sin q_2,
\nonumber
\eea
that close a $U_z(so(2,1))$ algebra under the canonical Poisson bracket
(\ref{sie}).

\sect{Casimirs and $N=2$ integrable systems}

Let us fix our attention on the examples of the previous section. If we
recall the deformed Casimir element (\ref{le}) and its non-deformed
counterpart
\be
C(J_1,J_2,J_3)=J_2^2 - J_1^2 - J_3^2,
\label{leclas}
\ee 
we know that both elements vanish, respectively, under the realizations
(\ref{hb}) and (\ref{ld}) (different canonical realizations will be
labelled by the real value obtained when the Casimir is represented).
However, if in the non-deformed $so(2,1)$ case we compute the coproduct of
the Casimir (\ref{leclas}), we get:
\bea
&& \co(C)=C(\co(J_1),\co(J_3),\co(J_2))\cr
&&\qquad =(1\otimes J_2 + J_2\otimes 1)^2 - (1\otimes J_1 + J_1\otimes
1)^2 - (1\otimes J_3 + J_3\otimes 1)^2\cr
&&\qquad =1\otimes C + C\otimes 1 + 2\,(J_2\otimes J_2 -J_1\otimes
J_1 -J_3\otimes J_3).
\label{hc}
\eea
When this abstract object is realized by using the $D$ representation we
obtain
\bea
&& C^{(2)}(q_1,q_2,p_1,p_2)\equiv (D\otimes D)(\co(C))=\cr
&& \qquad\qquad\quad= 0+0+2\,[p_1\, p_2 -(p_1\,\cos q_1)
\,(p_2\,\cos q_2) -(p_1\,\sin q_1)(p_2\,\sin q_2)]\cr
&&\qquad\qquad\quad=2\,p_1\,p_2\,(1- \cos(q_1-q_2)). 
\label{hd}
\eea
Therefore, although the Casimir vanishes on each space, the coproduct of $C$
has a ``crossed" contribution that is not trivial in the two-particle
realization.

This non trivial nature of $\co(C)$ is the cornerstone for the systematic
generation of a wide class of two dimensional integrable systems: in any
(Poisson) coalgebra endowed with a Casimir element, since the coproduct is
an algebra homomorphism and $C$ is a central element within $U(g)$, we can
conclude that \be
\pois{\co(C)}{\co(X_i)}_{A\otimes
A}=\co (\pois{C}{X_i}_A)=0,\qquad \forall X_i.
\label{sidc}
\ee
Consequently, if the Hamiltonian ${\cal{H}}(X_1,\dots,X_m)$ is an arbitrary
(smooth) function of the algebra generators we shall have that 
\be
\pois{\co(C)}{\co({\cal{H}}(X_1,\dots,X_m))}_{A\otimes
A}=\co (\pois{C}{{\cal{H}}(X_1,\dots,X_m)}_A)=0.
\label{sidh}
\ee

Therefore, a canonical realization of the coproduct of any (smooth)
function
$\cal{H}$ of the algebra generators of a coalgebra with Casimir element
$C$ defines a two-particle completely integrable Hamiltonian. In our case, 
any Hamiltonian
\bea
&& H^{(2)}(q_1,q_2,p_1,p_2):=(D\otimes
D)(\co({\cal{H}}(J_1,J_2,J_3)))=\cr
&& \qquad\qquad\quad=(D\otimes
D)({\cal{H}}(\co(J_1),\co(J_2),\co(J_3)))={\cal{H}}(f_1,f_2,f_3),
\label{hdos}
\eea
will always be in involution with the function $C(f_1,f_2,f_3)$ (\ref{hd}).
For instance, the function
\be
{\cal H}=J_2^2 + \k_2\,J_1^2 + \k_1\,J_3^2,
\label{ej1}
\ee
where $\k_1$ and $\k_2$ are real parameters (that have a precise
geometrical meaning in the context of pseudoorthogonal algebras
\cite{BHOS}), together with the $D$ realization and the formula
(\ref{hdos})
gives rise to the two-particle Hamiltonian \bea
&&\!\!\!\!\!\!\!\!\!\!  H^{(2)}(q_1,q_2,p_1,p_2)=(p_1 + p_2)^2 +
2\,p_1\,p_2(\k_2\,\cos q_1\,\cos q_2 + \k_1\,\sin q_1\,\sin q_2)\cr
&& \qquad\qquad\quad\quad + p_1^2(\k_2\,\cos^2 q_1 + \k_1\,\sin^2 q_1)
+ p_2^2(\k_2\,\cos^2 q_2 + \k_1\,\sin^2 q_2),
\label{ejham}
\eea
that defines a two-parameter family of integrable systems for which
(\ref{hd}) is a common constant of motion. If we specialize $\k_1=\k_2=1$,
we get
\be
H^{(2)}(q_1,q_2,p_1,p_2)=2(p_1^2 + p_2^2 + p_1\,p_2\,(1+
\cos(q_1-q_2)).
\label{ejkk}
\ee

At this point, some remarks are in order:

\noindent a) The choice of the Hamiltonian is constrained by
the requirement of functional independence between the two constants of the
motion. In particular, if we choose $\k_1=\k_2=-1$ we shall
recover the coproduct of the Casimir $2\,p_1\,p_2\,(1-
\cos(q_1-q_2))$, but now playing the role of the Hamiltonian. However,
integrability is now ensured by taking the coproduct of any generator as
the second constant of the motion (if $f_2$, we deduce the conservation of
the total momenta $p_1+p_2$). Note that, in general, the coproduct of a
given generator is not in involution with (\ref{ejham}).

\noindent b) Many different Hamiltonians may have the
same ``hidden" coalgebra symmetry, since different phase-space representations
and choices of the Hamiltonian function are possible.

\subsect{$N=2$ integrable deformations}

Now it is essential to stress that the integrable nature of this
construction is preserved for any possible coalgebra with Casimir element
that we could consider. Of course, deformations of Lie algebras with
coalgebra structure fall into this class and, therefore, can be used to
construct integrable systems.

Moreover:  if a Hamiltonian $H^{(2)}$ can be constructed by using the
previous procedure, any coalgebra deformation of its
symmetry algebra will generate an integrable deformation $H_z^{(2)}$ of
$H^{(2)}$ (provided that a deformed Casimir element $C_z$ and a deformed
canonical realization $D_z$ are available).

In particular, the standard quantum deformation of $so(2,1)$
(\ref{lb}--\ref{le}) can be used to define integrable two-particle
hamiltonians through the deformed coproduct of an arbitrary function
${\cal
H}$ of the generators:
\bea
&& H_z^{(2)}(q_1,q_2,p_1,p_2):=(D_z\otimes
D_z)(\Dco({\cal{H}}(\J_1,\J_2,\J_3)))=\cr
&& \qquad\qquad\quad=(D_z\otimes
D_z)({\cal{H}}(\Dco(\J_1),\Dco(\J_2),\Dco(\J_3)))=
{\cal{H}}(f_1^z,f_2^z,f_3^z).
\label{hdosz}
\eea
This deformed Hamiltonian will always be in involution with the
(deformed) phase space representation of the coproduct of the deformed
 Casimir
(\ref{le}), that reads:
\be
C_z^{(2)}(q_1,q_2,p_1,p_2)\equiv(D_z\otimes D_z)(\Dco(C_z))
= \pi_1\,\pi_2\,(1-\cos(q_1-q_2)), 
\label{lg}
\ee
where
\be
\pi_1=2\,\frac{\sinh (\tfrac{z}{2}
p_1)}{z}\,e^{\tfrac{z}{2}p_2},\qquad
\pi_2=2\,\frac{\sinh (\tfrac{z}{2}
p_2)}{z}\,e^{-\tfrac{z}{2}p_1}.
\label{lh}
\ee
An example of such a deformed Hamiltonian is provided by (\ref{ej1}) where
 the
generators are now replaced by their deformed counterparts
\be
{\cal H}=\J_2^2 + \k_2\,\J_1^2 + \k_1\,\J_3^2.
\label{defh}
\ee
From (\ref{hdosz}), and by making use of the deformed phase space
 realization
(\ref{dosq}), we get the integrable family of Hamiltonians
\bea
&& H_z^{(2)}(q_1,q_2,p_1,p_2)=
(f_2^z)^2 + \k_2\,(f_1^z)^2 + \k_1\,(f_3^z)^2 \cr
&&
\qquad\qquad  = (p_1 + p_2)^2 + 2\,\pi_1\,\pi_2\,( \k_2\,\cos q_1\,\cos q_2 
+ \k_1\,\sin q_1\,\sin q_2)\cr
&& 
\qquad\qquad\quad + \pi_1^2(\k_2\,\cos^2 q_1 +
\k_1\,\sin^2 q_1) + \pi_2^2(\k_2\,\cos^2 q_2 +
\k_1\,\sin^2 q_2).
\label{ejhamz}
\eea
Now, the deformation of the 
particular case $\k_1=\k_2=1$ reads
\be
H^{(2)}(q_1,q_2,p_1,p_2)=(p_1 + p_2)^2 + \pi_1^2 + \pi_2^2 +
2\,\pi_1\,\pi_2\,\cos(q_1-q_2).
\label{ejkkz}
\ee

Note that, after deformation, the case ($\k_1=\k_2=-1$) is no longer the
realization of the deformed Casimir (\ref{lg}). Of course, in order to get
the Casimir as a Hamiltonian we should consider ${\cal H}\equiv C_z$; then,
any $f_z^i$ can be taken as the remaining integral of the motion in
involution.  It also becomes apparent that the limit $z\to 0$ of
(\ref{ejhamz}) is just (\ref{ejham}).

\sect{Coassociativity and higher order coproducts}

The coassociativity constraint (\ref{co}) on $\co$ means that, in principle,
we could extend the previous procedure in order to get a more complex
system with three elementary constituents.  If we denote
$\co\equiv\co^{(2)}$ (in order to make more explicit the fact that
$\co$ defines a two--particle system) the mapping $\co^{(3)}:A\rightarrow
A\otimes A\otimes A$ has to be defined through (\ref{co}) by using one of
the following expressions:
\bea && \co^{(3)}:=(id\otimes\co^{(2)})\circ\co^{(2)},\cr  &&
\co^{(3)}:=(\co^{(2)}\otimes id)\circ\co^{(2)}. \label{fj}
\eea
From (\ref{co}), the result of this procedure is unique and does not depend
on the space within $A\otimes A$ we had chosen to duplicate.

On the other hand, it is well known that, once the coassociativity has
ensured the correctness of the three-constituents system, the construction
can be generalized to an arbitrary number of tensor products of $A$. For
instance, we would have that \be
\co^{(4)}:=(id\otimes id\otimes \co^{(2)})\circ\co^{(3)},
\label{fk}
\ee
will give rise to a fourth order coproduct starting from the third one.
In general, this procedure is described in the literature either by the
recurrence relation
\be
\co^{(N)}:=(id\otimes id\otimes\dots^{N-2)}\otimes id\otimes
\co^{(2)})\circ\co^{(N-1)},
\label{fl}
\ee
or by the following similar one 
\be
\co^{(N)}:=(\co^{(2)}\otimes id\otimes\dots^{N-2)}\otimes id\otimes
id)\circ\co^{(N-1)}.
\label{fla}
\ee
These definitions mean that given the $(N-1)$-th coproduct, the $N$-th one is
obtained by applying $\co^{(2)}$ onto the space located at the very right
(resp. left)  site. As a consequence, both (\ref{fl}) and (\ref{fla}) always
emphasize the role of such ``boundary" vector spaces within the tensor product.
This should not be necessary, since the essential meaning of coassociativity is
that all elementary spaces are equivalent in order to build up a larger
representation space by using the coproduct.

The algebraic transcription of this simple observation is the keystone
for all further developements included in this paper, and both the
recurrence character of (\ref{fl}) and its just mentioned ``asymmetry" can
be avoided by means of the following definition
\be
\co^{(N)}:=(\co^{(m)}\otimes \co^{(N-m)})\circ\co^{(2)},
\qquad \forall\,m=1,\dots,N-1,
\label{buena}
\ee
where $\co^{(1)}$ denotes the identity map $id$.
The proof of the equivalence between (\ref{buena}) and the usual one
(\ref{fl}) is given in the Appendix A. The meaning of this new
expression (\ref{buena}) can be made more clear with the use of Sweedler's
notation \cite{Sweed} that expresses the two-coproduct
$\co^{(2)}$ of an arbitrary element of the algebra as the linear
combination
\be
\co^{(2)}(X)=\sum_{\alpha}{X_{1\,\alpha}\otimes
X_{2\,\alpha} },
\label{Swee}
\ee
where $X_{1\,\alpha}$ and $X_{2\,\alpha}$ are functions depending on the
generators of the algebra.  By introducing this language in (\ref{buena})
we get that the the $N$-th coproduct of a generator reads
\be
\co^{(N)}(X):=\sum_{\alpha}{ \co^{(m)}(X_{1\,\alpha})\otimes
\co^{(N-m)}(X_{2\,\alpha}) }, \qquad \forall\,m=1,\dots,N-1,
\label{buena2}
\ee
which means that the final result can be obtained in $N-1$ different ways,
all of them equivalent, and given by the simultaneous application of two
lower degree coproducts on each of the two tensor components produced by
$\co^{(2)}$.

Now it is not difficult to prove, by induction, that $\co^{(N)}$ is an
algebra homomorphism between $A$ and ${A^{\otimes}}^N$
\be
\ponm{\co^{(N)}(X)}{\co^{(N)}(Y)}_{{A^{\otimes}}^N}=\co^{(N)}
(\ponm{X}{Y}_A),\qquad \forall X,Y\in A.
\label{fm}
\ee
A proof for this assertion can be found in the Appendix B. It is important
to stress that the symbol $\ponm{x}{y}$ denotes a general bracket, that can
be either the Poisson bracket for classical systems or the usual commutator
for quantum mechanical ones. The underlying algebraic structure is the same
for both kind of systems and the differences existing between them arise
from the different representation spaces we are working in. 

In particular,
in the case of $A\equiv U(g)$ and $\co^{(2)}$ given by (\ref{uea}), the
following $N$-th coproduct for the generators of $g$ is obtained:
\bea
&&\!\!\!\!\!\!\!\!\co^{(N)}(X_i)=X_i\otimes id\otimes id\otimes
\dots^{N-1)}\otimes id \cr
&&\qquad\qquad\quad + id\otimes X_i\otimes id\otimes\dots^{N-2)}\otimes
 id +
\dots \cr
&&\qquad\qquad\qquad\quad + id\otimes
id\otimes\dots^{N-1)}\otimes id\otimes X_i.
\label{fo}
\eea
which is just the definition of the usual ``total observable", and for
which the homomorphism condition is obviously fulfilled.

A more interesting example is provided by the deformation of (\ref{fo})
induced from (\ref{lb}). An iterative use of (\ref{fl}) leads to the
following expressions
\bea
&&\!\!\!\!\!\!\!\!\Dco^{(N)}(\J_2)=\J_2\otimes id\otimes id\otimes
\dots^{N-1)}\otimes
id \cr
&&\qquad\qquad\quad +
id\otimes \J_2\otimes id\otimes\dots^{N-2)}\otimes id + \dots \cr
&&\qquad\qquad\qquad\quad + id\otimes
id\otimes\dots^{N-1)}\otimes id\otimes \J_2,\cr
&&\!\!\!\!\!\!\!\!\Dco^{(N)}(\J_i)=\J_i\otimes e^{\tfrac{z}{2} \J_2}\otimes
e^{\tfrac{z}{2} \J_2}\otimes\dots^{N-1)}\otimes  e^{\tfrac{z}{2}
\J_2} \cr
&&\qquad\qquad\quad\quad + e^{-\tfrac{z}{2} \J_2}\otimes \J_i\otimes
e^{\tfrac{z}{2} \J_2}\otimes\dots^{N-2)}\otimes e^{\tfrac{z}{2} \J_2} 
+ \dots \cr
&&\qquad\qquad\qquad\quad\quad + e^{-\tfrac{z}{2} \J_2}\otimes
e^{-\tfrac{z}{2} \J_2}\otimes\dots^{N-1)}\otimes e^{-\tfrac{z}{2} \J_2}
\otimes \J_i,\qquad i=1,3.
\label{li}
\eea
Now, by taking into account that, if $X$ is a primitive generator and $h$
is an arbitrary complex parameter, the relation
$\co(e^{h\,X})=e^{h\,X}\otimes e^{h\,X}$ holds, we can choose any integer
$m$ running from 1 to $N-1$ and check that (\ref{li}) can be written in a
much more compact form
\bea
&&\!\!\!\!\!\! \Dco^{(N)}(\J_i)=\Dco^{(m)}(\J_i) \otimes
e^{\tfrac{z}{2} \J_2}\otimes\dots^{N-m)}\otimes e^{\tfrac{z}{2}
\J_2}+ e^{-\tfrac{z}{2} \J_2}\otimes\dots^{m)}\otimes
e^{-\tfrac{z}{2} \J_2}
\otimes \Dco^{(N-m)}(\J_i),\cr
&&\qquad\qquad = \Dco^{(m)}(\J_i) \otimes
e^{\tfrac{z}{2}\Dco^{(N-m)}(\J_2)} + 
e^{-\tfrac{z}{2}\Dco^{(m)}(\J_2)}\otimes \Dco^{(N-m)}(\J_i),
\label{lm}
\eea
that exactly corresponds to the result that we would have obtained by
directly applying (\ref{buena}). This expression was already used in
\cite{BCR} to demonstrate the integrability of a precise system constructed
from the standard deformation of $so(2,1)$.

\sect{The construction of $N$-particle Hamiltonians}

The procedure to obtain $N$=2 integrable systems presented in Section 3 can be
generalized to any number of degrees of freedom by making use of the
$N$-th coproduct. The statements here presented are valid for both
classical (Poisson) and quantum mechanical (commutator) realizations of
the underlying coalgebra $(A,\co)$. In order to emphasize this fact, the
symbol $\ponm{x}{y}$ will be used hereafter and the Appendix B contains the
computations that support this notation. On the other hand, the usual
embedding of $A\otimes A\otimes\dots^{m)}\otimes A$ within $A\otimes
A\otimes\dots^{N)}\otimes A$ as \be A\otimes A\otimes\dots^{m)}\otimes
A\otimes id \otimes\dots^{N-m)} \otimes id, \label{emb}
\ee
will be applied.

\subsect{General results}

The following Proposition holds:

\noindent $\bullet$ 
{\bf Proposition 1.} {\em Let $(A,\Delta)$ be a coalgebra with
generators
$X_i,\, i=1,\dots,l$ and Casimir element $C(X_1,\dots,X_l)$, and let us
consider the
$N$-th coproduct $\co^{(N)}(X_i)$ of the generators and the $m$-th
coproduct
 $\co^{(m)}(C)$ of the Casimir. Then,
\be
\ponm{\co^{(m)}(C)}{\co^{(N)}(X_i)}_{A\otimes
A\otimes\dots^{N)}\otimes A}=0,\qquad
i=1,\dots,l,\quad 1\leq m\leq N.
\label{za}
\ee}

\noindent PROOF: The case $m=N$ is easily proven by applying the
homomorphism property for the $N$-th coproduct. On the other hand, by
following Sweedler's notation, the 2nd coproduct of the Casimir can be
written as the sum
\be
\co^{(2)}(C)=\sum_{\alpha}{C_{1\,\alpha}\otimes C_{2\,\alpha} }.
\label{SweeC}
\ee
If we now compute (\ref{za}) we get 
\bea
&&\!\!\!\!\!\!\!\!\!\!\!\!\!\!\!\!\!\!\!\!\!\!
\ponm{\co^{(m)}(C)}{\co^{(N)}(X_i)}_{A\otimes\dots^{N)}\otimes A}=
\ponm{\co^{(m)}(C)\otimes
id\otimes\dots^{N-m)}\otimes id}{\co^{(N)}(X_i)}_{A\otimes\dots^{N)}
\otimes A}
\label{zb1}\\
&& \quad =\ponm{\co^{(m)}(C)\otimes id\otimes\dots^{N-m)}\otimes id}
{(\co^{(m)}\otimes \co^{(N-m)})\circ\co^{(2)}(X_i)}_{A\otimes\dots^{N)}
\otimes A}
\label{zb2}\\
&& \quad = \sum_{\alpha}{
\ponm{\co^{(m)}(C)\otimes id\otimes\dots^{N-m)}\otimes id}
{\co^{(m)}(C_{1\,\alpha})\otimes \co^{(N-m)}(C_{2\,\alpha})}_{A
\otimes\dots^{N)}
\otimes A} }\label{zb3}\\
&& \quad = \sum_{\alpha}{
\ponm{\co^{(m)}(C)}{\co^{(m)}(C_{1\,\alpha})}_{A\otimes\dots^{m)}\otimes A}
\otimes
\co^{(N-m)}(C_{2\,\alpha})}\label{zb4}\\
&& \quad = \sum_{\alpha}{
\co^{(m)}(\ponm{C}{C_{1\,\alpha}}_{A})\otimes
\co^{(N-m)}(C_{2\,\alpha}) }=0, \label{zb}
\eea
where (\ref{zb1}) reflects the usual embedding (\ref{emb}).
The next step (\ref{zb2}) includes the definition (\ref{buena}), that is
applied in (\ref{zb3}) with the help of (\ref{SweeC}). At this point the
identity functions in the first term allow us to split the
(Poisson/commutator) bracket as (\ref{zb4}), and the fact that we have
considered the $m$-th coproducts for the Casimirs leads to the final result
by taking into account that any order coproduct is a (Poisson/commutator)
map and that $\ponm{C}{C_{1\,\alpha}}_{A}=0$ for any
$C_{1\,\alpha}$
function.

This result provides a straightforward generalization of the $N=2$
construction of integrable systems sketched in Section 3:

\noindent $\bullet$ 
{\bf Theorem 2.} {\em Let $(A,\Delta)$ be a coalgebra with
generators $X_i,\, i=1,\dots,l$ and Casimir element $C(X_1,\dots,X_l)$ and
let ${\cal H}$ be an arbitrary (smooth/formal power series) function of the
generators of $A$. Then, the $N$-particle Hamiltonian
\be
H^{(N)}:=\co^{(N)}({\cal{H}}(X_1,\dots,X_l))=
{\cal{H}}(\co^{(N)}(X_1),\dots,\co^{(N)}(X_l)),
\label{htotg}
\ee
fulfills
\be
\ponm{C^{(m)}}{H^{(N)}}_{A\otimes
A\otimes\dots^{N)}\otimes A}=0,\qquad 1\leq m\leq N,
\label{za1}
\ee
where the $N$ Casimir elements $C^{(m)}$ ($m=1,\dots,N$)
are defined through
\be
C^{(m)}:= \co^{(m)}(C(X_1,\dots,X_l))=
C(\co^{(m)}(X_1),\dots,\co^{(m)}(X_l)).
\label{Ctotg}
\ee}

\noindent PROOF: The fact that $H^{(N)}$ and $C^{(N)}$ are in involution
is again a straightforward consequence of the homomorphism property of
$\co^{(N)}$. The rest of the proof follows directly from Proposition 1,
that tell us that the $N$-th coproduct of any generator commutes with all
the lower dimensional coproducts of the Casimir. Since our ${\cal H}$ is an
arbitrary function of such generators it will (Poisson)-commute with all the
$\co^{(m)}(C)$ elements.

\noindent $\bullet$ 
{\bf Corollary 3.} {\em In particular, all the $C^{(i)}$ elements
generated by the Casimirs are in involution
\be
\ponm{\co^{(k)}(C)}{\co^{(j)}(C)}=0,\qquad \forall\,k,j.
\label{cor3}
\ee}

To prove this assertion, it suffices to take $N=\mbox{max}\{k,j\}$
and apply the Theorem in the case ${\cal H}\equiv C$. This ensures the
involutivity among all the constants of motion. Note that, in principle, we
have a set of $N+1$ constants of motion
$\{C^{(1)},C^{(2)},\dots,C^{(N)},H^{(N)}\}$, but $C^{(1)}$ can be a real
number (see the examples in Section 3) and, in that case, we are left with
$N$ non-trivial integrals. On the other hand, functional independence among
them is guaranteed by the fact that each $C^{(i)}$ element lives on 
$A\otimes A\otimes\dots^{i)}\otimes A$ and that only 
$C^{(N)}$ and $H^{(N)}$ will share the same tensor space. In case $H^{(N)}$
is functionally dependent on $C^{(N)}$, we can always take the $N$-th
coproduct of any generator as the remaining independent constant of motion.

It is also immediate to check that, if our coalgebra has more than
one functionally independent Casimir elements $C_i$, the previous results
hold simultaneously for all of them.

\subsect{Classical mechanical systems}

The systematic construction of classical systems is provided by the
previous results when applied onto a Poisson coalgebra. Complete
integrability is obtained when a canonical realization $D$ of the Poisson
coalgebra is added to the general algebraic construction. As a
consequence, under such $D$, the Poisson bracket to be used is 
\be
\pois{f}{h}_{A\otimes A\otimes\dots^{N)}\otimes
A}=\sum_{i=1}^{N}{\left(\frac{\partial f}{\partial q_i} 
\frac{\partial h}{\partial
p_i} - \frac{\partial h}{\partial q_i} 
\frac{\partial f}{\partial p_i}\right)},
\label{fr}
\ee
the $N$-particle classical Hamiltonian is
written
\bea
&& \!\!\!\!\!\!\!\!\!\!\!\!\!\!\!\!
H^{(N)}(q_1,\dots,q_N,p_1,\dots,p_N):=(D\otimes\dots^{N)}\otimes D)
(\co^{(N)}({\cal{H}}(X_1,\dots,X_l)))=\cr
&& \qquad=(D\otimes\dots^{N)}\otimes D)
({\cal{H}}(\co^{(N)}(X_1),\dots,\co^{(N)}(X_l))) \cr
&& \qquad={\cal{H}}(
(D\otimes\dots^{N)}\otimes D)(\co^{(N)}(X_1)),\dots,
(D\otimes\dots^{N)}\otimes D)(\co^{(N)}(X_l))),
\label{htot}
\eea
and the $N-1$ Casimir functions $C^{(m)}$ ($m=1,\dots,N$)
read
\bea
&& \!\!\!\!\!\!\!\!\!\!\!\!\!\!\!\!
C^{(m)}(q_1,\dots,q_N,p_1,\dots,p_N):=(D\otimes\dots^{m)}\otimes D)
(\co^{(m)}(C(X_1,\dots,X_l)))=\cr
&& \qquad=(D\otimes\dots^{m)}\otimes D)
(C(\co^{(m)}(X_1),\dots,\co^{(m)}(X_l))) \cr
&& \qquad=C(
(D\otimes\dots^{m)}\otimes D)(\co^{(m)}(X_1)),\dots
(D\otimes\dots^{m)}\otimes D)(\co^{(m)}(X_l))).
\label{Ctot}
\eea
Since each space is linked to only
one degree of freedom, complete
integrability of the $N$-th Hamiltonian follows from (\ref{Ctot}).
If we are dealing with $r$ independent Casimir functions $C_i$,
the formalism can lead to the preservation of complete integrability for
any realization $D$ depending on $t$ pairs ($t\leq r$) of canonical
coordinates. Moreover,
nothing prevents us from the use of ``non-canonical" realizations, as we
shall see in the sequel. On the other hand, in case that other canonical
realizations $D',D'',\dots$ exist, their simultaneous use
in order to realize the tensor products of $A$ as, for instance,
$D\otimes D' \otimes D''\dots$ will provide ``mixed" realizations of the
same underlying abstract coalgebra. 

It is also important to stress
that no assumption concerning the explicit form of the coproduct is needed
to prove these statements. Therefore, deformed Poisson coalgebras can be
implemented with no difficulty within this algorithm in order to provide
(deformed) integrable systems, as it was done for $N=2$ in Section 3.1.

\subsect{Quantum mechanical systems}

Proofs of the aforementioned results when the commutator bracket is
considered offer no particular comments, up to the ones already included in
Appendix B, and the essential algebraic features of the general method here
presented are not modified by the non-commutativity of the algebra $A$ with
respect to the $(\cdot)$ product.

However, from a computational point of
view it is important to stress that in general extra contributions coming from
the unavoidable reordering processes will have to be considered.
Likewise, the quantum mechanical analogues of canonical realizations $D$
will be obtained either by using the generators $\hat p$ and $\hat
q$ of the Heisenberg-Weyl algebra or by means of
the so-called boson realizations in terms of the operators $a$ and
$a_+$ fulfilling $\conm{a}{a_+}=1$ (see \cite{BHN} and the references there
included for recent applications of bosonization procedures in the
representation theory of quantum algebras). Among the Hamiltonians that are
explicitly constructed in what follows, the ones expresed in terms of canonical
coordinates should be quantized in that way, and the remaining ones would lead
to quantum angular momentum (and, in particular, spin) chains.

\sect{Some coalgebra-invariant classical integrable systems}

We now present some examples of completely integrable systems obtained with
the aid of the previous results. Some of them are (to our
knowledge) new ones, and some others (although already known) are shown to
underly a ``hidden" coalgebra symmetry. Integrable deformations appear under
quantum coalgebra symmetry in a direct way.

\subsect{A $so(2,1)$ family including Calogero systems}

If we recall the (undeformed) $N$-th coproduct (\ref{fo}) for the $so(2,1)$
Poisson coalgebra and consider $N$ copies $D\otimes D\dots^{N)}\otimes D$
of the canonical phase space realization (\ref{hb}) we get the following
$N$-particle functions
\bea
&& f_1(\vec q,\vec p)=(D\otimes D\dots^{N)}\otimes
D)(\co^{(N)}(J_1))=\sum_{i=1}^{N}{p_i\,\cos q_i},\nonumber\\
&& f_2(\vec q,\vec p)=(D\otimes D\dots^{N)}\otimes
D)(\co^{(N)}(J_2))=\sum_{i=1}^{N}{p_i},\label{dosclasN}\\
&& f_3(\vec q,\vec p)=(D\otimes D\dots^{N)}\otimes
D)(\co^{(N)}(J_3))=\sum_{i=1}^{N}{p_i\,\sin q_i}.\nonumber
\eea
that close an $so(2,1)$ coalgebra. Now, if we take as Hamiltonian function
${\cal H}$ the quadratic two-parameter function (\ref{ej1}), Theorem 2
gives rise to the following integrable Hamiltonian
\be
H^{(N)}(\vec q,\vec p)=\left( \sum_{i=1}^{N}{p_i} \right)^2 +
\frac{1}{2}\sum_{i,j=1}^{N}{p_i\,p_j((\k_1+\k_2)\cos (q_i-q_j) -
(\k_1-\k_2)\cos (q_i+ q_j))}.
\label{qq1}
\ee
All the constants of motion in involution are given by the phase space
realizations of all the $m$-th coproducts of the Casimir function
$C=J_2^2 - J_1^2 - J_3^2$ (see (\ref{Ctot})), that read
\be
C^{(m)}(\vec q,\vec p) =\sum_{i<j}^{m}{2\,p_i\,p_j\,(1- \cos(q_i-q_j))},
\qquad m=2,\dots,N.
\label{he}
\ee
Note that, in the chosen realization, $C^{(1)}(q_1,p_1)=0$.

The case $\k_1=\k_2$ means that (\ref{qq1}) depends on the 
differences $(q_i-q_j)$. In particular, if we specialize the parameters
in the form $\k_1=\k_2=-1$, the chosen Hamiltonian coincides with the
Casimir. In that case, (\ref{qq1}) is just $C^{(N)}$ and (\ref{he}) gives us
$N-1$ constants of motion in involution (but -for instance- any $f_i$
function (\ref{dosclasN}) can be chosen to get a complete family of
integrals).  The system $H^{(N)}=\sum_{i<j}^{N}{2\,p_i\,p_j\,(1-
\cos(q_i-q_j))}$ was firstly introduced by Calogero \cite{Cal} as an
integrable Hamiltonian of the general type $H=\sum_{i<j}^{N}
{p_i\,p_j\,f(q_i-q_j)}$ (the Hamiltonian structures underlying an
integrable nonlinear shallow water equation with peaked solitons -the
so-called ``peakons" \cite{CH}- belongs to that class of systems). 

The ``hidden coalgebra symmetry" of this particular system was explicitly
introduced in \cite{BCR} and it was also implicitely stated in
\cite{KarAlg}. However, a crucial point is that any function $\cal{H}$ of
the generators (and not only the Casimirs) can be now taken as the
(integrable) Hamiltonian, thus generalizing the original Calogero model in a
highly arbitrary way. For instance, if we specialize the parameters as 
$\k_1=\k_2=1$, we arrive at the $N$-particle generalization of (\ref{ejkk}):
\be H^{(N)}(\vec q,\vec p)=\sum_{i=1}^{N}{2\, p_i^2} 
+ \sum_{i<j}^{N}{2\,p_i\,p_j(1 + \cos(q_i-q_j))},
\label{qq1N}
\ee
which is of course in involution with all the $C^{(m)}$ functions 
(\ref{he}).

\subsect{The algebra $h_4$ and an integrable oscillator chain }

Any other Lie algebra can give rise to an integrable system by following
the same procedure. For instance, we mention here the oscillator Lie algebra
$\osc$ is generated by
 $\{\aa,\ap,\am,\bb\}$  with Lie-Poisson brackets
\be
\pois{\aa}{\ap}=\ap,\quad \pois{\aa}{\am}=-\am,
\quad \pois{\am}{\ap}=\bb,\quad 
\pois{\bb}{\cdot}=0 .
\label{aa}
\ee
Besides the central generator $\bb$ there exists another Casimir
invariant for $\osc$: 
\be
C=\aa\bb - \ap\am.
\label{ab}
\ee 
A canonical $D$ realization for this algebra with vanishing  Casimir $C$ is
given by
\be
D(\aa)=p,\qquad
D(\ap)=\sqrt{p}\,e^{-q},\qquad
D(\am)=\sqrt{p}\,e^{q},\qquad
D(\bb)=1.
\label{hbos}
\ee

Let us now consider the ${\cal H}$ function
\be
{\cal H}=\lambda\,N + \mu\, \ap\,\am.
\label{hamosc}
\ee
It is immediate to check that, by using the primitive coproduct (\ref{fo})
for all the generators, Theorem 2 provides the following integrable
Hamiltonian
\be
 H^{(N)}(\vec q,\vec p)=(\lambda+\mu)\,\sum_{i=1}^{N}{p_i} 
+ 2\,\mu\,\sum_{i<j}^{N}{\sqrt{p_i\,p_j}\,\cosh (q_i- q_j)},
\label{qq2}
\ee
which is just the one introduced in \cite{CalOsc}. The integrals of the
motion in involution are given by the coproducts of the Casimir (\ref{ab})
in the chosen realization, and read: 
\be
C^{(m)}(\vec q,\vec p) =\sum_{i=1}^{m}{p_i} 
- \sum_{i<j}^{m}{2\,\sqrt{p_i\,p_j}\,\cosh (q_i- q_j)}.
\label{qq2c}
\label{heosc}
\ee
The quantization of the Hamiltonian (\ref{qq2}) has been performed in
 \cite{CalVDq}, where the equivalence between the quantum version of
(\ref{qq2}) and a system of coupled oscillators is shown (at this respect, 
see also \cite{KarOsc}).

\subsect{An integrable deformation from $U_z\,so(2,1)$}

The (standard) quantum deformation of $so(2,1)$ generates, through the
$N$-th order generalization of the comultiplication map (\ref{li}) and the
deformed realization $D_z$, an integrable deformation of the family
(\ref{qq1}). Let us fix $N$ and start by defining the quantities  
\be
\pi_k=2\,\frac{\sinh (\tfrac{z}{2}
p_k)}{z}\,\left(\prod_{i=1}^{k-1} e^{-\tfrac{z}{2}p_i} \right)
\,\left(\prod_{j=k+1}^{N} e^{\tfrac{z}{2}p_j} \right).
\label{lq}
\ee
The $N$-particle canonical (deformed) phase space realization will be
\bea
&& f_1^z(\vec q,\vec p)=(D_z\otimes D_z\dots^{N)}\otimes
D_z)(\co^{(N)}(\J_1))=\sum_{i=1}^{N}{\pi_i\,\cos q_i},\nonumber\\
&& f_2^z(\vec q,\vec p)=(D_z\otimes D_z\dots^{N)}\otimes
D_z)(\co^{(N)}(\J_2))=\sum_{i=1}^{N}{p_i},\label{doscuanN}\\
&& f_3^z(\vec q,\vec p)=(D_z\otimes D_z\dots^{N)}\otimes
D_z)(\co^{(N)}(\J_3))=\sum_{i=1}^{N}{\pi_i\,\sin q_i}.\nonumber
\eea
Now it is clear that, by taking as Hamiltonian function (\ref{defh}), the
Theorem 2 provides the following integrable Hamiltonian
\be
H^{(N)}(\vec q,\vec p)=\left( \sum_{i=1}^{N}{p_i} \right)^2 +
\frac{1}{2}\sum_{i,j=1}^{N}{\pi_i\,\pi_j((\k_1+\k_2)\cos (q_i- q_j) -
(\k_1-\k_2)\cos (q_i+ q_j))}.
\label{qq1z}
\ee

The integrals of the motion are just the $D_z$ realization of the $m$-th
deformed coproducts of $C_z$ (\ref{le}). A closed expression for them can
be readily obtained if we realize that the $\pi_i$ functions fulfill the
relation
\be
\tfrac{2}{z}\,\sinh (\tfrac{z}{2}
(p_1+p_2+\dots+p_m))=\pi_1 + \pi_2 + \dots + \pi_m.
\label{magia}
\ee
Now it is not difficult to check
that the explicit formula for $C_z^{(m)}$ is  \be
C_z^{(m)}=\sum_{i<j}^{m}  2\,\pi_i\,\pi_j\,(1- \cos(q_i-q_j)),
\qquad m=1,\dots,N
\label{lp}
\ee
where
and, as expected, in the limit $z\to 0$ we recover the ``classical"
Gaudin--Calogero system (\ref{he}).

Once again, the particular deformed system $\k_1=\k_2=-1$ does not coincide
with the
$N$-th Casimir function, although the former can be obtained from the
latter by substracting the function $\co^{(N)}
\left((2\,\frac{\sinh(\tfrac{z}{2}
\J_2)}{z})^2 - \J_2^2\right)$.

\subsect{A Ruijsenaars-Schneider-like model from a quantum deformation
of (1+1) Poincar\'e algebra}

The (1+1) Poincar\'e algebra ${\cal P}(1,1)$ is generated by
 $\{K,H,P\}$ and can be realized in Poisson form by the following brackets
\be
\pois{K}{H}=P,\quad \pois{K}{P}=H,\quad \pois{P}{H}=0.
\label{aap}
\ee
The known Casimir function for ${\cal P}(1,1)$ is: 
\be
C=H^2 - P^2,
\label{abp}
\ee 
and a $C=1$ Poisson realization of this algebra in terms of a canonical
coordinate $q$ and its conjugate rapidity $\t$ is the following:
\be
D(K)=q,\qquad
D(H)=\cosh\t,\qquad
D(P)=\sinh\t.
\label{hbosRS}
\ee
If we consider the primitive coproduct (\ref{fo}) and take as hamiltonian
function just the $H$ generator, the resultant coalgebra-induced integrable
system reads:
\be
 H^{(N)}(\vec q,\vec \t)=\sum_{i=1}^{N}{\cosh \t_i},\qquad 
C^{(m)}(\vec q,\vec \t) =m + \sum_{i<j}^{m}{2\,\cosh(\t_i- \t_j)}.
\label{nond}
\ee
Note that the associated dynamics is quite trivial since (\ref{nond}) 
depends only on the canonical momenta.

However, a completely different system is derived when we consider the
(non-coboundary) quantum deformation $U_z{\cal P}(1,1)$ given by the
deformed coproduct
\bea
&& \Dco(K) =1 \otimes K + K\otimes 1,\cr  
&&  \Dco(H) =e^{-\tfrac{z}{2}K} \otimes H + H\otimes
e^{\tfrac{z}{2}K}; \label{lbp} \\
&&  \Dco(P) =e^{-\tfrac{z}{2}K} \otimes P + P\otimes
e^{\tfrac{z}{2}K},\nonumber
\eea
that, in spite of the non-triviality of the deformation, is still
compatible with the undeformed brackets (\ref{aap}). This deformation was
firstly introduced in $\cite{VK}$, and it was later recognized as the dual
of Woronowicz's quantum (pseudo)Euclidean group \cite{Wo}.

Therefore, the compatibility with (\ref{aap}) implies that the phase space
realization (\ref{hbosRS}) is also valid in the deformed case. If we 
consider again the time translation $H$ as the  hamiltonian function {\cal
H}, the  $N$-th generalization of the deformed coproduct and the phase
space realization (\ref{hbosRS}) gives rise to the integrable system
defined by
\be
H_z^{(N)}(\vec q,\vec \t)=\sum_{i=1}^{N}{\cosh \t_i 
\, \exp{\left ( -\tfrac{z}{2}(\sum_{j=1}^{i-1}{q_j}) 
+\tfrac{z}{2}(\sum_{k=i+1}^{N}{q_k}) \right ) } }.
\label{qq2RS}
\ee
This system presents strong analogies with respect to the so-called
Ruijsenaars-Schneider Hamiltonian \cite{RS}, which is a relativistic
analogue of Calogero-Moser systems. 

The integrals of motion are obtained, as usual, from the $N$-th order
deformed coproducts of the Casimir (\ref{abp}). A straightforward
computation shows that they are
\be
C_z^{(m)}(\vec q,\vec \t) =\sum_{i<j}^{m}{2\,\cosh(\t_i- \t_j)
\, \exp{\left(-\tfrac{z}{2}\, (q_i - q_j) - z\,(\sum_{l=1}^{i-1}{q_l})
+ z\, (\sum_{k=j+1}^{N}{q_k}) \right)}  }.
\label{caspo}
\ee
Note that in this case additional integrals appear due to the fact that $P$
commutes with $H$. In particular, the deformed $N$-th coproduct of $P$
\be
P_z^{(N)}(\vec q,\vec \t)=\sum_{i=1}^{N}{\sinh \t_i 
\, \exp{\left ( -\tfrac{z}{2}(\sum_{j=1}^{i-1}{q_j}) 
+\tfrac{z}{2}(\sum_{k=i+1}^{N}{q_k}) \right ) } }.
\label{qq2RSP}
\ee
will Poisson-commute with both $H_z^{(N)}$ and $C_z^{(N)}$.

\sect{Angular momentum realizations}

The coalgebra symmetry that gives rise to $N$ integrals of motion in
involution is not restricted to the use of canonical realizations. We shall
consider in this Section its application to the construction of classical
integrable ``angular momentum" chains through the $so(2,1)$ Poisson coalgebra
given by a primitive coproduct and the Poisson bivector
\be
\Lambda=\s_3 \, \partial_{\s_2}\wedge\partial_{\s_1} - 
\s_1 \, \partial_{\s_2}\wedge\partial_{\s_3} + 
\s_2 \, \partial_{\s_3}\wedge\partial_{\s_1} ,
\label{biqc}
\ee
afterwards, its deformed counterpart (\ref{biq}) will be examined and the
consequences of the deformation analysed.

These examples will also stress the possibilities of applying the actual
formalism to the quantum mechanical context. From the following examples it
will become clear that quantization will imply  (up to
sometimes important contributions coming from reordering) the
substitution of the $\sigma$ coordinates by the corresponding Pauli
matrices. In this way, the $so(2,1)$ systems can be interpreted as Gaudin
magnets, and the quantum deformation of the coalgebra will introduce a
variable range interaction in the model. An exhaustive study of these
aspects will be presented elsewhere.

\subsect{The $so(2,1)$ model: classical XYZ Gaudin magnet}

Let us now consider the Poisson bracket (\ref{biqc})
corresponding to the $so(2,1)$ Lie algebra, which is tantamount to consider
-in our language- the realization $S$ 
\be
S(J_2)=\s_2,\qquad
S(J_1)=\s_1,\qquad
S(J_3)=\s_3,
\label{hbs}
\ee
that will be completely defined provided the value $c=\s_2^2 - \s_1^2 -
\s_3^2$ is given. Now, a straightforward replica of the generalized
Calogero systems of the previous section is provided by $N$-copies of
(\ref{hbs}) (that we shall distinguish with the aid of a superindex
$\s_i^k$
and that could have different values $c_k$ of the Casimir) and  the
(undeformed) $N$-th coproduct (\ref{fo}). Therefore, we have the (quite
trivial) realization for the coproducts \be (S\otimes S\dots^{N)}\otimes
S)(\co^{(N)}(\s_i))=\sum_{k=1}^{N}{\s_i^k}, \qquad i=1,2,3.
\label{con}
\ee
If we preserve (\ref{ej1}) as Hamiltonian function, the Theorem 2
provides the following integrable Hamiltonian:
\bea
&&\!\!\!\!\!\!\!\!\!\!\!\!\!\!\!  H^{(N)}(\vec{\s})=
\left( \sum_{l=1}^{N}{\s_2^l} \right)^2 + 
\k_2\, \left( \sum_{l=1}^{N}{\s_1^l} \right)^2 +
\k_1\, \left( \sum_{l=1}^{N}{\s_3^l} \right)^2 \cr
&&
=\sum_{l=1}^{N}{\left\{ (\s_2^l)^2 + \k_2\, (\s_1^l)^2 + \k_1\,
(\s_3^l)^2      \right\}} + 
2\,\sum_{i<j}^{N}{\left\{ \s_2^i\,\s_2^j +
\k_2\,\s_1^i\,\s_1^j  + \k_1\,\s_3^i\,\s_3^j\right\} }.
\label{gau}
\eea
That is, a classical long range interacting XYZ angular momentum chain of
the Gaudin type \cite{Gau,GHik,BGHP}.   

The constants of motion are derived from the $m$-th coproducts of the
Casimir function $C=J_2^2 - J_1^2 - J_3^2$ in the usual way and read:
\be
C^{(m)}(\vec{\s}) =
\sum_{l=1}^{m}{c_l} + 
2\,\sum_{i<j}^{m}{\s_2^i\,\s_2^j - \,\s_1^i\,\s_1^j 
-\,\s_3^i\,\s_3^j}.
\label{gauc}
\ee
Since the first term is constant, we are lead to the hyperbolic XXX-Gaudin
system. Note that this system becomes the keystone for the integrability
of any finite chain obtained through an arbitrary function of the $so(2,1)$
generators. On the other hand, this construction can be immediately
quantized by transforming (\ref{hbs}) into a representation in terms
of angular momentum operators and by taking into account the corresponding
discrete values for the Casimir operators.

\subsect{$U_z\,(so(2,1))$ and XYZ model with variable range exchange}

Let us now construct an integrable deformation of the XYZ classical Gaudin
system through $U_z\,(so(2,1))$. The Poisson realization $S_z$ that we
are going to consider is
\be
S_z(\J_2)=\j_2,\qquad
S_z(\J_1)=\j_1,\qquad
S_z(\J_3)=\j_3,
\label{hbsz}
\ee
with $c_z$ given by (\ref{sym}). Note that the $\j_i$ coordinates are not
the classical ones (they live on a deformed hyperboloid (\ref{sym})),
although we shall consider a particular representation in terms of the
classical structure (\ref{biqc}) later.

As usual, the comultiplication map (\ref{li}) and the
chosen realization $S_z$ gives rise to the following
functions expressing the $N$-th order coproduct of the $\J_i$ generators:
\bea
&& (S_z\otimes S_z\dots^{N)}\otimes
S_z)(\co^{(N)}(\J_1))=\sum_{l=1}^{N}{\left(\prod_{i=1}^{l-1} 
e^{-\tfrac{z}{2}\j_2^i} \right)
\,\j_1^l\,\left(\prod_{j=l+1}^{N} e^{\tfrac{z}{2}\j_2^j} \right)},
\nonumber\\
&& (S_z\otimes S_z\dots^{N)}\otimes
S_z)(\co^{(N)}(\J_2))=\sum_{l=1}^{N}{\j_2^l},\label{cops}\\
&& (S_z\otimes S_z\dots^{N)}\otimes
S_z)(\co^{(N)}(\J_3))=\sum_{l=1}^{N}{\left(\prod_{i=1}^{l-1} 
e^{-\tfrac{z}{2}\j_2^i} \right)
\,\j_3^l\,\left(\prod_{j=l+1}^{N} e^{\tfrac{z}{2}\j_2^j} \right)}.\nonumber
\eea
If we consider now the Hamiltonian function (\ref{defh}), its $N$-th order
coproduct leads, through the usual method, to the following deformation of
the clasical Gaudin XYZ system (\ref{gau}):
\be
H_z^{(N)}(\vec{\j})=
\sum_{l=1}^{N}{\left\{ (\j_2^l)^2 + e^{2z\beta_i}(\k_2 (\j_1^l)^2 + 
\k_1 (\j_3^l)^2)      \right\}} + 
2\sum_{i<j}^{N}{\left\{ \j_2^i\j_2^j +
e^{z\alpha_{ij}}(\k_2\j_1^i\j_1^j  + \k_1\j_3^i\j_3^j)\right\} },
\label{gaud}
\ee
where the $\beta,\alpha$ functions depend on $\j_2$ as follows:
\bea
&& \beta_i= -\tfrac{1}{2}(\sum_{j=1}^{i-1}{\j_2^j}) 
+\tfrac{1}{2}(\sum_{k=i+1}^{N}{\j_2^k}),\cr
&& \alpha_{ij}=\beta_i + \beta_j= -\tfrac{1}{2}\, (\j_2^i - \j_2^j)
- \sum_{l=1}^{i-1}{\j_2^l} + \sum_{k=j+1}^{N}{\j_2^k}.
\label{ij}
\eea

This Hamiltonian corresponds to a sort of XYZ Gaudin magnet with variable
range anisotropy given by the $\alpha_{ij}$ functions. In the limit  $z\to
0$ we recover the non deformed XYZ system (\ref{gau}). Note that the
commutativity among the $\j_i^l$ allows such a compact final expression,
that will certainly contain additional terms in the quantum mechanical
case. The complete integrability of such Hamiltonian is ensured by the
$m$-th deformed coproducts $(m\leq N)$ of $C_z$ (\ref{le}) in the $S_z$
representation. A closed expression for them is not difficult to find by
recalling the formula (\ref{magia}): \be C_z^{(m)}(\vec{\j})=
\sum_{l=1}^{m}{ e^{2\,z\,\beta_i}\,C_z^i      } + 
2\,\sum_{i<j}^{m}{e^{z\,\alpha_{ij}}\,\left\{ \frac{\sinh (\tfrac{z}{2}
\j_2^i)}{z/2}\,\frac{\sinh (\tfrac{z}{2}
\j_2^j)}{z/2} - \j_1^i\,\j_1^j  -\j_3^i\,\j_3^j\right\} },
\label{sscas}
\ee
where $C_z^i$ are the corresponding deformed Casimir functions on each
lattice site. As usual, the $N$-th Casimir can be considered as the
Hamiltonian. In that case, any of the coproducts (\ref{cops}) can be used
to complete the integrals of the motion.

\subsubsect{The zero representation}

We insist now in the fact that the $\j_i$ coordinates are deformed ones.
However, realizations in terms of the non deformed variables $\s_j$ are
available. In particular, let us consider the (deformed) Poisson realization
$U_z$ \be
U_z(\J_2)=\s_2,\qquad
U_z(\J_1)=\frac{\sinh (\tfrac{z}{2}
\s_2)}{\s_2\,z/2}\s_1,\qquad
U_z(\J_3)=\frac{\sinh (\tfrac{z}{2}
\s_2)}{\s_2\,z/2}\s_3.
\label{hbsu}
\ee
The functions defined by (\ref{hbsu}) close an
$U_z\,so(2,1)$ under the Poisson bracket (\ref{biqc}) and provided that
the classical coordinates are defined on the $c=0$ cone $\s_2^2 - \s_1^2 -
\s_3^2=0$. In this case, the previous construction leads to the following
Hamiltonian: \bea
&&\!\!\!\!\!\!\!\!\!\!\!\!\!\!\!  H_z^{(N)}(\vec{\s})=
\sum_{l=1}^{N}{\left\{ (\s_2^l)^2 + e^{2\,z\,\beta_i}\,\left(\frac{\sinh
(\tfrac{z}{2} \s_2^l)}{\s_2^l\,z/2}\right)^2\,(\k_2\, (\s_1^l)^2 + 
\k_1\,(\s_3^l)^2)      \right\}}   \cr 
&&\qquad\quad
+ 2\,\sum_{i<j}^{N}{\left\{ \s_2^i\,\s_2^j +
e^{z\,\alpha_{ij}}\,\frac{\sinh (\tfrac{z}{2}
\s_2^i)}{\s_2^i\,z/2}\,\frac{\sinh (\tfrac{z}{2}
\s_2^j)}{\s_2^j\,z/2}\,(\k_2\,\s_1^i\,\s_1^j  + \k_1\,\s_3^i\,\s_3^j)
\right\}
}.
\label{gauds}
\eea
The constants of motion are easily computed and read (in our representation
$C_z=0$):
\be
C_z^{(m)}(\vec{\s})= 
2\,\sum_{i<j}^{m}{e^{z\,\alpha_{ij}}\,\frac{\sinh (\tfrac{z}{2}
\s_2^i)}{\s_2^i\,z/2}\,\frac{\sinh (\tfrac{z}{2}
\s_2^j)}{\s_2^j\,z/2}\,\left\{ \s_2^i\,\s_2^j - \s_1^i\,\s_1^j 
-\s_3^i\,\s_3^j\right\} }.
\label{sss}
\ee
In this case, $C_z^{(m)}$ are hyperbolic Gaudin hamiltonians with variable
range exchange. An analysis of long-range Hamiltonians and some examples of
variable range interacting systems can be found in \cite{BGHP}
and \cite{Inoz}, respectively.

\sect{Concluding remarks}

Summarizing, we have demonstrated that any algebra $A$ endowed with a
coassociative coproduct $\co$ (either deformed or not) can be seen as the
abstract object that, after choosing a given representation, gives rise in a
direct and systematic way to a wide class of $N$-dimensional integrable
systems (with $N$ finite but arbitrary). Within this class of systems, the
original coalgebra is not only a set of symmetries, but the algebraic
object that generates explicitly the Hamiltonian and the constants of
motion. Moreover, the theory can be used to generate both classical and
quantum systems by choosing, respectively, either a Poisson or an
operatorial realization of $A$.

The universality of the coalgebra-induced construction that we have
presented in this paper suggests a number of further investigations in
different contexts. From a general point of view we would like to mention
the  unsolved question concerning the existence of a Lax formulation for
this scheme and its connection with the integrability properties of the
known quantum algebra invariant Hamiltonians. On the other hand, a symmetry
method in order to decide whether a known system is coalgebra invariant or
not would be evidently helpful. In this sense, the long-range interacting
nature of our construction is worth to be emphasized, although it could not
be essential (we recall that known quantum algebra invariant systems
usually contain only nearest neighbour interactions).

As a consequence arising at a purely ``classical" level, phase space
realizations $D$ of Lie algebras become relevant tools in order to
construct new examples. If such a realization exists in terms of only one
pair of canonical coordinates, complete integrability is ensured. However,
for Lie algebras with rank greater than one, both the existence of various
Casimir functions and the possibility of having $D$ realizations depending
on more than one canonical pair have to be taken into account in order to
analyse the complete integrability of the system. 

The explicit solutions for the examples here
presented deserve further investigations aswell. Known results concern the
$so(2,1)$ Calogero system defined through the Casimirs (\ref{he}), that was
already solved in \cite{Cal}. The $N=2$ deformed motion has been also
shown to be solvable (and it includes a deformed period) in \cite{Cor}. For
arbitrary $N$, the quantum deformation can be seen as a displacement from 
the geodesic motion (on a proper
manifold) that characterizes the non-deformed system. All these results
concerning the canonical realization should be completed and translated
into the behaviour of the Gaudin systems defined through the angular
momentum Poisson bracket.

Finally, we think that these results provide a strong physical motivation
for Hopf algebra deformations, since now they
could be sistematically used to generate new integrable systems (we recall
that the (1+1) Poincar\'e example shows that such deformed systems can be
interesting even when their non-deformed counterparts are associated to
trivial dynamics). It is known that the number of Hopf algebra deformations
for a given $U(g)$ is not arbitrary: in fact, their classification is
intimately linked to the notion of Lie bialgebra and, for some low
dimensional cases, complete (and constructive) classifications of quantum
deformations have been recently obtained \cite{Osc}. Therefore, a coalgebra
invariant Hamiltonian constructed from a given $g$ can be
``integrably" deformed in a finite number of ways that, at least in some
cases, can be explicitly obtained and will certainly provide a better
understanding of the physical relevance of coalgebra symmetries.

\bigskip
\bigskip

\noindent
{\Large{{\bf Acknowledgments}}}

\bigskip

\noindent A.B. has been partially supported by DGICYT (Project  PB94--1115)
from the Ministerio de Educaci\'on y Ciencia de Espa\~na and by Junta de
Castilla y Le\'on (Projects CO1/396 and CO2/297) and acknowledges the
hospitality during his stays in Rome; O.R has been partially supported by
I.N.F.N., Sezione di Roma,  and is grateful  to  the Physics Department of
Burgos University for its hospitality.   
  The authors  would like to thank Francisco
Herranz for a careful reading of the manuscript.


\bigskip
\bigskip

\noindent
{\Large{{\bf Appendix A}}}
\bigskip

\appendix

\setcounter{equation}{0}

\renewcommand{\theequation}{A.\arabic{equation}}

\noindent The equivalence between the definition (\ref{buena}) and
the usual ones (\ref{fl}-\ref{fla}) is obvious for the $N=3$ case, being
expressed in terms of the coassociativity condition (\ref{fj}).
The case $N=4$ is also easy to check by direct computation. Therefore, we
shall prove the general case by induction, by taking into account that, for
a generic $N+1$, we have to prove that any value of $m=1,\dots,N$ in the
definition (\ref{buena}) leads to (\ref{fl}-\ref{fla}).

We shall assume that 
\be
\co^{(N)}:=(\co^{(m)}\otimes \co^{(N-m)})\circ\co^{(2)},
\qquad \forall\,m=1,\dots,N-1,
\label{a1}
\ee
holds, and we have to prove that
\be
\co^{(N+1)}:=(\co^{(k)}\otimes \co^{(N-k+1)})\circ\co^{(2)},
\qquad \forall\,k=1,\dots,N.
\label{a2}
\ee

If we denote $id^{(r)}\equiv id\otimes id\dots^{r)}\otimes id$, from
(\ref{fla}) we can compute $\co^{(N+1)}$ in the following way: \bea
&&\co^{(N+1)}=(\co^{(2)}\otimes id^{(N-1)}) \circ \co^{(N)} \cr
&&\qquad\qquad\!\!\!=(\co^{(2)}\otimes id^{(N-1)}) \circ 
(\co^{(m)}\otimes \co^{(N-m)}) \circ \co^{(2)} \cr
&& \qquad\qquad\!\!\! = ( ((\co^{(2)}\otimes id^{(m-1)}) \circ \co^{(m)}
)    \otimes \co^{(N-m)}  ) \circ \co^{(2)}  \cr
&& \qquad\qquad\!\!\! = (\co^{(m+1)}\otimes \co^{(N-m)}) \circ \co^{(2)}.
\label{a3}
\eea
where we can choose $\forall\,m=1,\dots,N-1$. Therefore, the validity of
(\ref{a2}) is proven for $k=2,\dots,N$.

The only relation which remains to be proven is the case $k=1$, that reads
\be
\co^{(N+1)}:=(id\otimes \co^{(N)})\circ\co^{(2)}.
\label{a4}
\ee
In this case we can compute its equivalence with respect to the known
recurrence (\ref{fla}) as follows:
\bea
&&\co^{(N+1)}=(\co^{(2)}\otimes id^{(N-1)}) \circ \co^{(N)} \cr
&&\qquad\qquad\!\!\!=(\co^{(2)}\otimes id\otimes id^{(N-2)}) \circ
\co^{(N)} \cr
&&\qquad\qquad\!\!\!=(\co^{(2)}\otimes id\otimes id^{(N-2)}) \circ 
(\co^{(2)}\otimes \co^{(N-2)}) \circ \co^{(2)} \cr
&& \qquad\qquad\!\!\! = ( ((\co^{(2)}\otimes id) \circ \co^{(2)} ) 
\otimes \co^{(N-2)}  ) \circ \co^{(2)}  \cr
&& \qquad\qquad\!\!\! = ( ((id\otimes \co^{(2)}) \circ \co^{(2)} ) 
\otimes \co^{(N-2)}  ) \circ \co^{(2)}  \cr
&&\qquad\qquad\!\!\!=(id\otimes \co^{(2)}\otimes id^{(N-2)}) \circ 
(\co^{(2)}\otimes \co^{(N-2)}) \circ \co^{(2)} \cr
&&\qquad\qquad\!\!\!=(id\otimes \co^{(2)}\otimes id^{(N-2)}) \circ 
\co^{(N)} \cr
&&\qquad\qquad\!\!\!=(id\otimes \co^{(2)}\otimes id^{(N-2)}) \circ 
(id\otimes \co^{(N-1)}) \circ \co^{(2)} \cr
&&\qquad\qquad\!\!\!=(id\otimes (( \co^{(2)}\otimes id^{(N-2)}) \circ
 \co^{(N-1)}) )\circ  \co^{(2)} \cr
&&\qquad\qquad\!\!\!=(id\otimes \co^{(N)})\circ\co^{(2)}.
\label{a5}
\eea
Finally, note that the equivalence between the ordinary definitions
(\ref{fl}) and (\ref{fla}) is obtained as a byproduct from this derivation
by considering the $k=N$ case.

\newpage


\bigskip
\bigskip

\noindent
{\Large{{\bf Appendix B}}}
\bigskip
\appendix

\setcounter{equation}{0}

\renewcommand{\theequation}{B.\arabic{equation}}

\noindent The aim of this second Appendix is to prove the homomorphism
condition (\ref{fm}) that we shall split into the commutator and Poisson
cases, respectively. As usual, the $N=2$ case is part of the definition of a
(Poisson) Hopf algebra, and we shall proceed by induction.

\medskip

\noindent $\bullet$ {\bf a) $\co^{(N)}$ as a homomorphism.}

\noindent Let us consider the algebra $A$ endowed with an associative
product $(\cdot)$ that now we shall explicitly write. We know that, by
definition, the coproduct $\co^{(2)}$ is a homomorphism between $A$ and
$A\otimes A$: \be
\co^{(2)}(X \cdot Y) = \co^{(2)}(X) \cdot \co^{(2)}(Y),
\qquad\forall\,X,Y\,\in A.
\label{b0}
\ee
If we assume that $\co^{(N-1)}$ is a homomorphism,
by using Sweedler's notation,
\be
\co^{(2)}(X)=\sum_{\alpha}{X_{1\,\alpha}\otimes
X_{2\,\alpha} }, \qquad\quad
\co^{(2)}(Y)=\sum_{\beta}{Y_{1\,\beta}\otimes
Y_{2\,\beta} }.
\label{b3}
\ee
and by recalling the definition of $\co^{(N)}$ in terms of $\co^{(N-1)}$ and
$\co^{(2)}$, we have that
\bea
&& \!\!\!\!\!\!\!\!\!\!\!\! \co^{(N)}(X) \cdot \co^{(N)}(Y) = 
((\co^{(N-1)}\otimes id) \circ \co^{(2)}(X)) \cdot 
((\co^{(N-1)}\otimes id) \circ \co^{(2)}(Y)) \cr
&&\qquad = \sum_{\alpha,\beta} { \left( \co^{(N-1)}(X_{1\,\alpha}) \otimes
X_{2\,\alpha} \right) \cdot \left( \co^{(N-1)}(Y_{1\,\beta}) \otimes
Y_{2\,\beta} \right) }\cr
&&\qquad = \sum_{\alpha,\beta} { \co^{(N-1)}(X_{1\,\alpha} \cdot
Y_{1\,\beta}
) \otimes X_{2\,\alpha} \cdot Y_{2\,\beta} }\cr
&&\qquad = (\co^{(N-1)}\otimes id) \left( \sum_{\alpha,\beta}  {
X_{1\,\alpha} \cdot Y_{1\,\beta}  \otimes X_{2\,\alpha} \cdot
Y_{2\,\beta} } \right) \cr
&&\qquad = (\co^{(N-1)}\otimes id) \circ \co^{(2)}(X \cdot Y) =
\co^{(N)}(X \cdot Y).
\label{b02}
\eea
This result holds for $(\cdot)$ being either a commutative or a
non-commutative product. In the latter case, the homomorphism condition for
the commutator $\conm{X}{Y}:=X \cdot Y - Y \cdot X$ is immediately deduced
from this result.

\medskip
\noindent $\bullet$ {\bf b) $\co^{(N)}$ as a Poisson map.}

\noindent Let us assume that $(A,\co^{(2)})$ is a Poisson-Hopf algebra and
that the $(N-1)$-th coproduct fulfills
\be
\pois{\co^{(N-1)}(X)}{\co^{(N-1)}(Y)}_{A\otimes
A\otimes\dots^{N-1)}\otimes A}=\co^{(N-1)} (\pois{X}{Y}_A),\qquad
\forall X,Y\in
A.
\label{b1}
\ee
(Hereafter we shall supress the subscripts that label the space where the
Poisson bracket is defined). From (\ref{buena}) we can write
\be
\pois{\co^{(N)}(X)}{\co^{(N)}(Y)}=
\pois{(\co^{(N-1)}\otimes id) \circ \co^{(2)}(X) }
{(\co^{(N-1)}\otimes id) \circ \co^{(2)}(Y) }.
\label{b2}
\ee
With the aid of (\ref{b3}) we can compute it explicitly:
\bea
&&\!\!\!\!\!\!\!\!\! \pois{\co^{(N)}(X)}{\co^{(N)}(Y)} \cr
&& =
\pois{(\co^{(N-1)}\otimes id)
\big( \sum_{\alpha}{X_{1\,\alpha}\otimes
X_{2\,\alpha} }\big) }
{(\co^{(N-1)}\otimes id)
\big( \sum_{\beta}{Y_{1\,\beta}\otimes
Y_{2\,\beta} } \big) } \cr
&& =\sum_{\alpha,\beta}{ \pois{ \co^{(N-1)}(X_{1\,\alpha})\otimes
X_{2\,\alpha} }{ \co^{(N-1)}(Y_{1\,\beta})\otimes
Y_{2\,\beta}} } \cr
&& = \sum_{\alpha,\beta} \big( \pois{ \co^{(N-1)}(X_{1\,\alpha})}
{\co^{(N-1)}(Y_{1\,\beta})} \otimes X_{2\,\alpha}\cdot Y_{2\,\beta} \cr
&& \qquad\qquad +
(\co^{(N-1)}(X_{1\,\alpha})\cdot \co^{(N-1)}(Y_{1\,\beta})) \otimes 
\pois{ X_{2\,\alpha}}{Y_{2\,\beta}} \big) \cr
&& = \sum_{\alpha,\beta}{ \left( \co^{(N-1)}\big(
\pois{X_{1\,\alpha}}{Y_{1\,\beta}}\big) \otimes
(X_{2\,\alpha}\cdot Y_{2\,\beta})  +
\co^{(N-1)}(X_{1\,\alpha}\cdot Y_{1\,\beta}) \otimes 
\pois{ X_{2\,\alpha}}{Y_{2\,\beta}} \right) }\cr
&& = \sum_{\alpha,\beta}{ (\co^{(N-1)}\otimes id)\left(
\pois{X_{1\,\alpha}}{Y_{1\,\beta}} \otimes (X_{2\,\alpha}\cdot
Y_{2\,\beta}) + (X_{1\,\alpha}\cdot Y_{1\,\beta}) \otimes \pois{
X_{2\,\alpha}}{Y_{2\,\beta}} \right) } \cr
&& = \sum_{\alpha,\beta}{ (\co^{(N-1)}\otimes id)\left(
\pois{X_{1\,\alpha} \otimes X_{2\,\alpha}}{Y_{1\,\beta} \otimes
Y_{2\,\beta}} \right) } \cr
&& = (\co^{(N-1)}\otimes id)\left( \pois{\sum_{\alpha}{X_{1\,\alpha}\otimes
X_{2\,\alpha} }}{\sum_{\beta}{Y_{1\,\beta}\otimes
Y_{2\,\beta} }} \right) \cr
&& = ((\co^{(N-1)}\otimes id) \circ \co^{(2)})(\pois{X}{Y})=\co^{(N)}
(\pois{X}{Y}).
\label{b4}
\eea
Throughout this computation we have used the Poisson-map condition
(\ref{b1}) for $\co^{(N-1)}$ and the homomorphism condition for the
(now commutative) $(\cdot)$ product in the Poisson algebra. Note that the proof
b) of the commutative (Poisson) case is more involved. In this classical
mechanical context we have to impose the compatibility of $\co$ with
respect to {\it two} independent products: the (commutative) ``pointwise" one
$(\cdot)$ and the Poisson bracket $\pois{\,}{\,}$. On the contrary, in the
``quantum-mechanical" case the latter is replaced by the commutator, which is
constructed in terms of the former (now non-commutative).




\begin{thebibliography}{40}



\bibitem{KBI} Korepin V E, Bogoliubov N M and Izergin A G 1993 {\it
Quantum Inverse Scattering Method and Correlation Functions}, Cambridge
University Press 

\bibitem{GRAS} G\'omez C, Ruiz-Altaba M and Sierra G 1995 {\it
Quantum Groups in Two-Dimensional Physics}, Cambridge University Press 

\bibitem{FRT} Reshetikhin N Y, Takhtadzhyan L A and Fadeev L D 1990 {\it
Leningrad. Math. J.} {\bf 1}  193

\bibitem{Dr} Drinfel'd V G 1986
{\it Quantum Groups}, Proceedings of the International Congress of
Mathematics,  MRSI Berkeley, p.798

\bibitem{PS} Pasquier V and Saleur H 1990 {\it Nucl. Phys.} {\bf B330} 523

\bibitem{KS} Kulish P P and Sklyanin E K 1991 {\it J. Phys. A} {\bf 24} 
L435

\bibitem{ACF} Arnaudon D, Chryssomalakos C and Frappat L 1995
{\it J. Math. Phys.} {\bf 36} 5262

\bibitem{Chang} Chang Z 1995 {\it Phys. Rep.} {\bf 262}  137 

\bibitem{Foerster} Foerster A  1996 {\it J. Phys. A} {\bf 29} 7625

\bibitem{CP} Chari V and Pressley A 1994 {\it A Guide to Quantum Groups},
Cambridge University Press 

\bibitem{Gerst} Gerstenhaber M 1964 {\it Ann. Math.} {\bf 79}  59\\
Gerstenhaber M and Schack S D 1992 {\it Contemp. Math.} {\bf
134}  51

\bibitem{BFFLS} Bayen F, Flato M, Fronsdal C, Lichnerowicz A and
Sternheimer D 1978 {\it Ann. Phys.} {\bf 111}  61

\bibitem{Hopf} Hopf H 1941 {\it Ann. Math.} {\bf 42}  22

\bibitem{CGST} Celeghini E, Giachetti R, Sorace E and Tarlini M 1992
{\it Contractions of Quantum Groups}, Lecture Notes in Mathematics, n.1510,
p.221\\
Lukierski J, Ruegg H, Nowicky A and Tolstoy V N
1991 {\it Phys. Lett.} {\bf B264}  331\\
Ballesteros A, Gromov N A, Herranz F J, del Olmo M and Santander M 
1995 {\it J. Math. Phys} {\bf 36} 5916

\bibitem{Osc} Ballesteros A and Herranz F J 1996 {\it J. Phys. A} {\bf 29}
4307\\
Ballesteros A, Herranz F J and Parashar P 1997 {\it J. Phys. A} {\bf 30}
L149

\bibitem{OP} Olshanetski M A and Perelomov A M 1981 {\it Phys. Rep.} {\bf
71}  313

\bibitem{Per} Perelomov A M 1990 {\it Integrable Systems of Classical
Mechanics and Lie algebras}, Birkh\"auser

\bibitem{BCR} Ballesteros A, Corsetti M and Ragnisco O 1996 {\it
Czech. J. Phys} {\bf 46} 1153; Proceedings of the V International 
Colloquium ``Quantum Groups and Integrable Systems", Prague, June 1996 

\bibitem{KarAlg} Karimipour V 1996 {\it Algebraic and geometric structure
of the integrable models recently proposed by Calogero} hep-th/9602161\\
{\it Relation of the new Calogero models
and XXZ spin chains} hep-th/9604092

\bibitem{Cal} Calogero F 1995 {\it Phys. Lett.} {\bf A201} 306

\bibitem{CalOsc} Calogero F 1995 {\it J. Math. Phys.} {\bf 36} 9

\bibitem{VK} Vaksman L and Korogodskii L I 1989
{\it Sov. Math. Dokl.} {\bf 39} 173

\bibitem{RS} Ruijsenaars S N M and Schneider H  
1986 {\it Ann. Phys.} {\bf 170} 370\\
Nijhoff F W, Kuznetsov V B, Sklyanin E K and Ragnisco O 1996 {\it J. Phys. A}
{\bf 29} L333

\bibitem{Gau} Gaudin M 1983 {\it La Fonction d'Onde de Bethe}, Masson
(Paris)

\bibitem{Inoz} Frahm H and Inozemtsev V I 
1994 {\it J. Phys. A} {\bf 27} L801; cond-mat/9512071

\bibitem{Post} Postnikov M 1982 {\it Lie Groups and Lie Algebras}, Mir
Publishers, Moscow

\bibitem{Ji} Jimbo M 1985 {\it Lett. Math. Phys} {\bf 10}  63

\bibitem{BHOS} Ballesteros A, Herranz F J, del Olmo M and Santander M 
1993 {\it J. Phys. A} {\bf 26} 5801

\bibitem{Sweed} Sweedler M E 1969 {\it Hopf algebras},
Benjamin, New York 

\bibitem{BHN} Ballesteros A, Herranz F J and Negro J
1997 {\it J. Phys. A} {\bf 30} 6797

\bibitem{CH} Camassa R and Holm D D 1993 {\it Phys. Rev. Lett.} {\bf 71} 
1661

\bibitem{CalVDq} Calogero F and van Diejen J F 1995 {\it Phys. Lett.} 
{\bf A205} 143

\bibitem{KarOsc} Karimipour V 1997 {\it J. Math. Phys.} {\bf 38} 1577

\bibitem{Wo}  Woronowicz S L 1992
{\it Commun. Math. Phys.} {\bf 149} 637\\
Ballesteros A, Celeghini E, Giachetti R, Sorace E and Tarlini M 1993 {\it J.
Phys. A} {\bf 26} 7495

\bibitem{GHik} Hikami K
1995 {\it J. Phys. A} {\bf 28} 4997

\bibitem{BGHP} Bernard D, Gaudin M, Haldane F N M and Pasquier V
1993 {\it J. Phys. A} {\bf 26} 5219

\bibitem{Cor} Corsetti M 1996 MSc Thesis, Universit\'a La Sapienza, Roma. 





















\end{thebibliography}
\end{document}